\documentclass[11pt]{article}
\usepackage[utf8]{inputenc}
\usepackage[T1]{fontenc}
\usepackage{arxiv}          
\usepackage{amsmath}
\usepackage{amsfonts}
\usepackage{amssymb}
\usepackage[numbers,super,sort&compress]{natbib} 
\usepackage{graphicx}
\usepackage[colorlinks,citecolor=blue,urlcolor=blue,linkcolor=blue]{hyperref}
\usepackage{url}
\usepackage{booktabs}
\usepackage{microtype}
\usepackage{doi}
\usepackage[figuresright]{rotating}
\usepackage{multirow}
\usepackage{array}
\usepackage{colortbl}       
\usepackage{caption}
\usepackage{subfig}         
\usepackage{xcolor}
\usepackage{makecell}

\newcommand{\btheta}{\boldsymbol{\theta}}
\newcommand{\bSigma}{\boldsymbol{\Sigma}}
\newcommand{\bD}{\mathbf{D}}

\graphicspath{{./images/}}

\title{Parametric inference for the average hazard approach in time-to-event analysis with sparse risk sets}
\renewcommand{\shorttitle}{Parametric inference for the average hazard approach with sparse risk sets}
\renewcommand{\headeright}{Qian, Tian, and Uno}
\renewcommand{\undertitle}{}

\date{}

\author{
    {Zihan~Qian} \\
    Department of Biostatistics \\
    Massachusetts General Hospital \\
    Boston, Massachusetts, USA
    \And
    {Lu~Tian} \\
    Stanford University \\
    Stanford, California, USA
    \And
    {Hajime~Uno}\thanks{Corresponding author: huno@ds.dfci.harvard.edu} \\
    Department of Data Science \\
    Dana-Farber Cancer Institute \\
    Boston, Massachusetts 02215, USA
}

\begin{document}
\maketitle

\baselineskip=12pt

\begin{abstract}
The average hazard (AH) is a recently proposed model-free summary of an event-time distribution. Over a prespecified window $[0,\tau]$, it equals the cumulative incidence probability at $\tau$ divided by the restricted mean survival time at $\tau$ and can be interpreted as a person-time incidence rate that does not depend on the censoring distribution. 
The difference and ratio of AHs from two groups, DAH and RAH, are between-group contrasts on absolute and relative scales without requiring proportional hazards. 
Existing nonparametric inference for AH and its contrasts is based on the Kaplan--Meier estimator of the survival function, and can become unreliable when risk sets become sparse or exhausted near $\tau$, as may occur in stratified analyses. 
As a practical remedy, we investigate parametric AH inference using Weibull and generalized gamma models. 
We use maximum-likelihood estimators and corresponding confidence intervals for group-specific AHs, DAH and RAH in both unstratified and stratified analyses. Simulations show that under correct model specification, little bias and near-nominal coverage can be achieved in settings where nonparametric inference is inadequately supported by observed data. 
We illustrate the method using a randomized trial comparing neratinib plus capecitabine with lapatinib plus capecitabine in metastatic breast cancer. 
Parametric inference provides a practical complement to nonparametric AH analysis when sparse or exhausted risk sets make inference at clinically relevant truncation times less reliable.
\end{abstract}


\section{Introduction}
\label{sec:intro}
In the analysis of time-to-event outcomes, the Cox proportional hazards (PH) model\cite{Cox:1972un} has long served as the predominant method for comparing two treatment groups. Its semiparametric formulation and close connection to the log-rank test have made it a convenient statistical tool in clinical and epidemiological research. However, as has been widely discussed,\cite{Uno:2014ii,Chappell:2016dm} the Cox hazard ratio (HR) has important limitations as a summary of the treatment effect. When the PH assumption is violated, the HR estimator converges to a quantity that depends on the censoring distribution, which, in turn, is affected by the accrual and follow-up patterns of the study. This quantity is therefore not determined solely by the event-time distributions, which are of our interest. Moreover, without a reference level, it is difficult to interpret the magnitude of HR as a relative measure in clinical practice. These limitations have motivated the development of new summary measures that are defined directly from the event-time distributions and remain interpretable without requiring PH assumption.

One such measure is the \textit{average hazard with survival weight} (AH), a group-specific summary of the event-time distribution over a prespecified time window.\cite{Uno2023-sm}
The same quantity has also been referred to as the \textit{generalized hazard} in the recent literature.\cite{Snapinn2022-wf,Jackson2025-vq}
Specifically, the AH of an event time $T$ is
\[
\eta(\tau)
=
\frac{1-S(\tau)}{\int_0^{\tau} S(u)\,du}
=
\frac{E\{I(T\leq\tau)\}}{E(T\wedge\tau)},
\]
where $S(t)=\Pr(T>t)$ and $\tau$ is a truncation time selected a priori.  That is, the AH is the cumulative incidence probability at $\tau$, $F(\tau)=1-S(\tau)$, divided by the restricted mean survival time (RMST) at $\tau$, $R(\tau)=\int_0^{\tau}S(u)\,du$.
It can be interpreted as a person-time incidence rate over $[0,\tau]$: the expected number of events per unit of expected event-free time accumulated by $\tau$.
Correspondingly, when $F(\tau)>0$, the reciprocal $1/\eta(\tau)=R(\tau)/F(\tau)$  can be interpreted as the expected event-free time accumulated per event by $\tau.$

Two properties are particularly relevant to the present work. First, the AH is censoring-free in the sense that its population value is determined solely by the event-time distribution and does not involve the censoring distribution. By contrast, the conventional incidence rate, the number of observed events divided by the observed person-time\cite{rothman2008modern}, converges to a quantity that depends on the censoring-time distribution unless the hazard is constant.\cite{Uno2023-sm}
Second, the AH is a group-specific absolute summary. Estimating the AH separately in each group provides a reference value in control group to facilitate the interpretation of DAH and RAH and is consistent with recommendations to report treatment effects in both absolute and relative terms.\cite{annals,Hopewell2025-we}

Nonparametric inference procedures for group-specific AHs and the corresponding DAH and RAH have been developed using Kaplan--Meier estimators.\cite{Uno2023-sm} Extensions to stratified analyses and regression settings have also been proposed.\cite{Uno2024-rg,Qian2025-bd} An important practical question, however, is how to conduct reliable inference for AH when too few patients remain under follow-up near $\tau$, the end of the time window. When the risk set is too small near $\tau$, the large-sample normal approximation to relevant statistics becomes inaccurate, and the actual coverage probability of resulting confidence intervals can be quite different from the nominal level (Section~\ref{sec:simulation}). When no patient remains at risk at $\tau$, the point estimate of AH itself may not be computable. This problem is especially acute in stratified analyses, because AH needs to be estimated by treatment group within each stratum.\cite{Qian2025-bd} This practical problem is the focus of the present paper.

\begin{figure}[!tbp]
\centering
\subfloat[(A) Hormone receptor--negative disease]{\includegraphics[width=\linewidth]{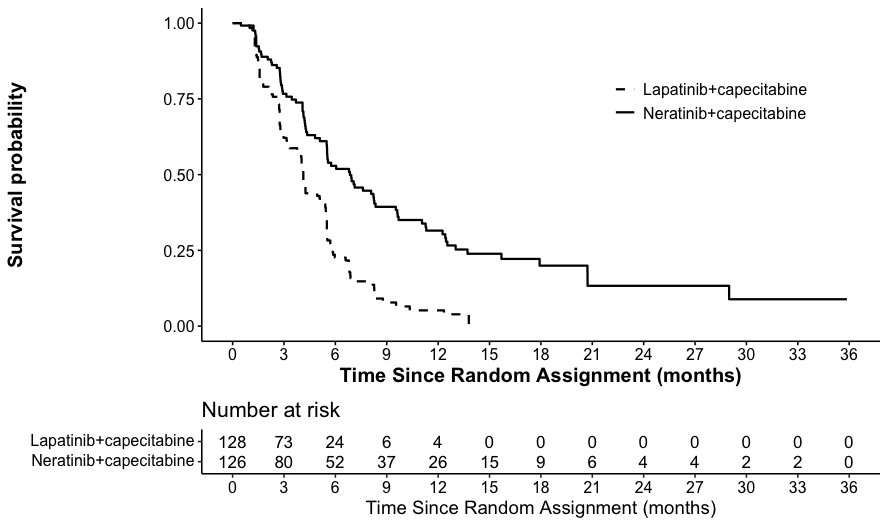}}\\
\subfloat[(B) Hormone receptor--positive disease]{\includegraphics[width=\linewidth]{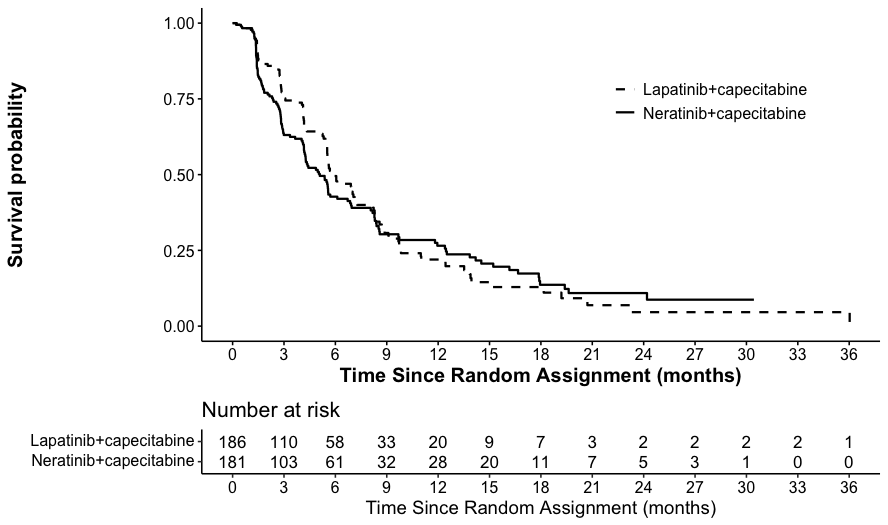}}
\caption{
Kaplan--Meier curves for progression-free survival (PFS) by treatment group within strata defined by hormone receptor status: (A) hormone receptor--negative disease and (B) hormone receptor--positive disease. Patient-level data were reconstructed from the curves published by Saura et al.\cite{saura2020neratinib} using the method of Guyot et al.\cite{guyot2012enhanced} N+C, neratinib plus capecitabine; L+C, lapatinib plus capecitabine. \label{fig:KM}
}
\end{figure}

A motivating example comes from a randomized phase~III trial comparing neratinib plus capecitabine (N+C) with lapatinib plus capecitabine (L+C) in patients with HER2-positive metastatic breast cancer.\cite{saura2020neratinib} Progression-free survival (PFS) was a co-primary endpoint, and hormone receptor status was used as a stratification factor in randomization. We reconstructed patient-level PFS data from the published Kaplan--Meier curves using the method of Guyot et al.\cite{guyot2012enhanced} Figure~\ref{fig:KM} shows the Kaplan--Meier curves by treatment group within each stratum based on the reconstructed data. In the hormone receptor--positive stratum, the two curves crossed at approximately eight months, suggesting nonproportional hazards. Of more direct relevance to the present paper, PFS events occurred rapidly in the L+C group of the hormone receptor--negative stratum. Among 128 patients, only four remained at risk at month 12, whereas follow-up extended well beyond two years in the other treatment-by-stratum combinations. Consequently, the L+C group of the hormone receptor--negative stratum severely limits the choice of truncation time for the stratified analysis as the relevant survival function is only estimable up to approximately 12 months, despite substantially longer follow-up in the other groups (Section~\ref{sec:example}). This example illustrates how limited follow-up in a single group/stratum can restrict AH-based inference for an otherwise well-followed study.

To address this issue, we propose a parametric procedure for inference on AH that complements, rather than replaces, the established nonparametric procedure. In unstratified analyses, we fit a parametric survival model, such as a Weibull or generalized gamma model, separately to each treatment group. In stratified analyses, we fit a separate model to each treatment group within each stratum. The fitted survival functions yield simple plug-in estimates of treatment-specific AH with corresponding confidence intervals (CIs). This approach permits inference at clinically relevant truncation times for which Kaplan--Meier-based inference is not adequately supported because the relevant risk sets have been exhausted. In such settings, inference necessarily relies on model-based extrapolation, and its validity depends on the adequacy of the assumed parametric models. The aim is therefore practical: to preserve a clinically meaningful analysis window, when the nonparametric procedure would otherwise require an impractically early truncation time $\tau$ or fail to support reliable inference.

The remainder of this paper is organized as follows. Section~\ref{sec:methods} describes parametric estimation of the group-specific AH, DAH, and RAH and the associated delta-method inference for both unstratified analyses and stratified analyses based on the standardized AH. Section~\ref{sec:example} illustrates the approach using the motivating breast cancer data, including model diagnostics. Section~\ref{sec:simulation} reports numerical studies, conducted under correct model specification, that illustrate how nonparametric inference becomes overly conservative and then inadequately supported as risk sets become sparse or exhausted, and that evaluate the finite-sample bias and coverage of the parametric procedure in those settings. These studies are not intended to assess robustness to model misspecification. Section~\ref{sec:discussion} discusses practical implications, limitations, and future work. Technical details, diagnostics for the example, and supplementary simulation results are provided in Appendices~\ref{app:parm-details}--\ref{app:sim}.

\section{Methods}
\label{sec:methods}
\subsection{Notation, estimands, and assumptions}
\label{ssec:notation}

Let $T_j$ and $C_j$ denote the event time and the right-censoring time in group $j\in\{0,1\}$, where $j=1$ denotes the treatment group and $j=0$ the control group, and let $(T_{ji},C_{ji})$, $i=1,\ldots,n_j$, be independent copies of $(T_j,C_j)$. The observed data in group $j$ are $\{(X_{ji},\Delta_{ji});\, i=1,\ldots,n_j\}$, where $X_{ji}=\min(T_{ji},C_{ji})$ and $\Delta_{ji}=I(T_{ji}\le C_{ji})$. We assume a parametric model for the event-time distribution in group $j$, indexed by $\btheta_j\in\Theta_j\subset\mathbb{R}^{d_j}$, with survival function $S_j(t;\btheta_j)$, density $f_j(t;\btheta_j)$, and hazard $h_j(t;\btheta_j)=f_j(t;\btheta_j)/S_j(t; \btheta_j)$.

For a truncation time $\tau>0$, the AH in group $j$ is
\begin{equation}
\label{eq:AH}
\eta_j(\tau;\btheta_j)
= \frac{\int_0^{\tau} h_j(u;\btheta_j) S_j(u;\btheta_j)\,du}
       {\int_0^{\tau} S_j(u;\btheta_j)\,du}
= \frac{1 - S_j(\tau;\btheta_j)}{R_j(\tau;\btheta_j)},
\end{equation}
where $R_j(\tau;\btheta_j)=\int_0^{\tau} S_j(u;\btheta_j)\,du$ is the RMST at $\tau$. The first expression shows that the AH is a weighted average of the hazard function over $[0,\tau]$, with weights proportional to the survival function, which gives the measure its name. The between-group contrasts are
$\mathrm{DAH}(\tau)=\eta_1(\tau;\btheta_1)-\eta_0(\tau;\btheta_0)$ and
$\mathrm{RAH}(\tau)=\eta_1(\tau;\btheta_1)/\eta_0(\tau;\btheta_0)$.

Because the inference below is model-based, we state the underlying assumptions together for clarity. These assumptions apply separately to each treatment group in unstratified analyses and to each treatment group within each stratum in stratified analyses (Section~\ref{ssec:strat}).
\begin{description}
\item[(A1)] The censoring time is independent of the event time.
\item[(A2)] Observations are independent and identically distributed.
\item[(A3)] The assumed parametric family contains the true event-time distribution.
\item[(A4)] The usual regularity conditions for maximum likelihood hold: the parameter is identifiable, its true value lies in the interior of the parameter space, and the information matrix is nonsingular.
\item[(A5)] With generic notations, $S(t;\btheta)$ and $\partial S(t;\btheta)/\partial\btheta$ are continuous in $(t,\btheta)$ on $[0,\tau]\times\mathcal{N}$, where $\mathcal{N}$ is a compact set containing the true parameter value, and $\eta(\tau;\btheta)>0$.
\item[(A6)] In a stratified analysis, the number of strata $K$ is fixed, $n_{jk}/n\to p_{jk}>0$ for treatment group $j$ within stratum $k$, as the total sample size $n \to \infty$; the standardization weights are fixed (Section~\ref{ssec:strat}).
\end{description}
The validity of the inference procedure relies on assumption (A3). If the assumed parametric model is misspecified, the fitted model does not converge to the true event-time distribution. Instead, it converges to the member of the assumed model family that provides the best approximation to the true distribution in terms of the Kullback--Leibler divergence, often referred to as the pseudo-true model \cite{White1982}. Consequently, inference on the AH based on the fitted model may be misleading under model misspecification. Model assessment and sensitivity analyses are discussed in Section~\ref{ssec:modelchoice}.

\subsection{Estimation and inference in unstratified analyses}
\label{ssec:unstrat}

The log-likelihood for $\btheta_j$ based on the right-censored data in group $j$ is
\begin{equation}
\label{eq:loglik}
\ell_j(\btheta_j)
= \sum_{i=1}^{n_j} \left\{ \Delta_{ji} \log f_j(X_{ji};\btheta_j)
+ (1-\Delta_{ji}) \log S_j(X_{ji};\btheta_j) \right\}.
\end{equation}
The model may be fitted with all observed data or with the modified data administratively censored at $\tau$; in the latter case, $(X_{ji},\Delta_{ji})$ in \eqref{eq:loglik} is replaced by $(X_{ji}^{(\tau)},\Delta_{ji}^{(\tau)})$, where $X_{ji}^{(\tau)}=\min(X_{ji},\tau)$ and $\Delta_{ji}^{(\tau)}=\Delta_{ji}I(X_{ji}\le\tau)$. Under correct model specification, the inference below applies to either fit; the choice between them, including its implications under model misspecification, is discussed in Section~\ref{ssec:modelchoice}.

Let $\widehat{\btheta}_j$ denote the maximum likelihood estimator (MLE) and
$\widehat{\mathbf{I}}_j = -\partial^2 \ell_j(\btheta_j)/\partial \btheta_j \partial \btheta_j^\top\big|_{\btheta_j=\widehat{\btheta}_j}$
the observed information matrix. Under (A1)--(A6), $\widehat{\btheta}_j$ is consistent and asymptotically normal with a covariance matrix $\bSigma_j$, which can be  estimated by $\widehat{\boldsymbol{\Sigma}}_j=\widehat{\mathbf{I}}_j^{-1}$.

The plug-in estimator of the group-specific AH is
\begin{equation*}
\widehat{\eta}_j(\tau)=\eta_j\!\left(\tau;\widehat{\btheta}_j\right)
= \frac{1 - S_j\!\left(\tau;\widehat{\btheta}_j\right)}
       {R_j\!\left(\tau;\widehat{\btheta}_j\right)},
\end{equation*}
and the estimated contrasts are
$$\widehat{\mathrm{DAH}}(\tau)=\widehat{\eta}_1(\tau)-\widehat{\eta}_0(\tau)~~\mbox{and}~~\widehat{\mathrm{RAH}}(\tau)=\widehat{\eta}_1(\tau)/\widehat{\eta}_0(\tau).$$

Variances of AH, DAH and RAH estimates can be derived using the delta method. Specifically, let
\[\bD_j(t; \btheta_j)=\frac{\partial F_j(t; \btheta_j)}{\partial \btheta_j} ~~\mbox{and}~~ F_j(t; \btheta_j)=1-S_j(t; \btheta_j).\]
Then, the gradient of \eqref{eq:AH} with respect to $\btheta_j$ can be written as
\begin{equation}
\label{eq:grad_eta}
\nabla_{\btheta_j}\, \eta_j(\tau;\btheta_j)
= \frac{\bD_{j}(\tau;\btheta_j)}{R_{j}(\tau; \btheta_j)}
      + F_{j}(\tau; \btheta_j) \int_0^\tau \frac{\bD_{j}(t;\btheta_j)dt}
       {R_{j}(\tau; \btheta_j)^{2}},
\end{equation}
and the estimated variance of $\hat{\eta}_j(\tau)$ is
\begin{equation}
\label{eq:var_eta}
\widehat{\mathrm{Var}}\!\left\{\widehat{\eta}_j(\tau)\right\}
=
\big[\nabla_{\btheta_j}\, \eta_j(\tau;\widehat{\btheta}_j)\big]^{\!\top}
\, \widehat{\boldsymbol{\Sigma}}_j \,
\big[\nabla_{\btheta_j}\, \eta_j(\tau;\widehat{\btheta}_j)\big].
\end{equation}
As observations from two treatment groups are independent, the variance of $\widehat{\mathrm{DAH}}$ and $\log(\widehat{\mathrm{RAH}})$ can be estimated by
\begin{align}
\label{eq:var_DAH}
\widehat{\sigma}^2_D(\tau)
&= \widehat{\mathrm{Var}}\!\left\{\widehat{\eta}_1(\tau)\right\}
       + \widehat{\mathrm{Var}}\!\left\{\widehat{\eta}_0(\tau)\right\},\\
\label{eq:var_logRAH}
\mbox{and}~~\widehat{\sigma}^2_{R}(\tau)
&= \frac{\widehat{\mathrm{Var}}\!\left\{\widehat{\eta}_1(\tau)\right\}}{\widehat{\eta}_1(\tau)^2}
       + \frac{\widehat{\mathrm{Var}}\!\left\{\widehat{\eta}_0(\tau)\right\}}{\widehat{\eta}_0(\tau)^2},
\end{align}
respectively. 

Wald-type CIs follow directly. For the AH in group $j,$ we may construct a CI for $\log\eta_j(\tau)$ first and then convert it to that for $\eta_j(\tau).$ The resulting $(1-\alpha)$CI for $\eta_j(\tau)$ is
\[
\left[\widehat{\eta}_j(\tau)\exp\!\left\{- z_{1-\alpha/2}\,\frac{\sqrt{\widehat{\mathrm{Var}}\{\widehat{\eta}_j(\tau)\}}}{\widehat{\eta}_j(\tau)}\right\}, ~~~\widehat{\eta}_j(\tau)\exp\!\left\{ z_{1-\alpha/2}\,\frac{\sqrt{\widehat{\mathrm{Var}}\{\widehat{\eta}_j(\tau)\}}}{\widehat{\eta}_j(\tau)}\right\} \right],\]
where $z_{1-\alpha/2}$ denotes the $(1-\alpha/2)$ quantile of the standard normal distribution. Similarly, the $(1-\alpha)$CI for RAH can be constructed as
\[
\left[\widehat{\mathrm{RAH}}(\tau)\exp\!\left\{-z_{1-\alpha/2}\widehat{\sigma}_R(\tau)\right\},~~~\widehat{\mathrm{RAH}}(\tau)\exp\!\left\{z_{1-\alpha/2}\widehat{\sigma}_R(\tau)\right\}\right].
\]
The CI for DAH can be constructed directly using (\ref{eq:var_DAH}):
\[\left[\widehat{\mathrm{DAH}}(\tau)-z_{1-\alpha/2}\widehat{\sigma}_D(\tau),~~~ \widehat{\mathrm{DAH}}(\tau)+z_{1-\alpha/2}\widehat{\sigma}_D(\tau)\right].\] 

The null hypotheses $H_0: \eta_1(\tau)=\eta_0(\tau)$ can be tested with either Wald statistic
$$Z_{\mathrm{DAH}}= \frac{\widehat{\mathrm{DAH}}(\tau)}{\widehat{\sigma}_D(\tau)}$$ 
or
$$Z_{\mathrm{RAH}}=\frac{\log\widehat{\mathrm{RAH}}(\tau)}{\widehat{\sigma}_R(\tau)},$$
which is asymptotically standard normal under the null hypothesis.

\subsection{Stratified analysis via direct standardization}
\label{ssec:strat}

The standardized AH estimand used here is that proposed by Qian et al.,\cite{Qian2025-bd} where the standardized survival distribution as a weighted mixture of the stratum-specific survival distributions and the standardized AH as the AH of that mixture distribution. We retain the same estimand and replace the stratum-specific Kaplan--Meier estimators with fitted parametric survival functions, yielding parametric plug-in estimators and the corresponding likelihood-based inference.

Suppose that there are $K$ strata. Let $n_{jk}$ denote the sample size of group $j$ in stratum $k$.  The total sample size $n=\sum_{j=0}^{1}\sum_{k=1}^{K}n_{jk}$, and let $S_{jk}(t;\btheta_{jk})$ be the corresponding survival function and $R_{jk}(\tau;\btheta_{jk})=\int_0^{\tau}S_{jk}(u;\btheta_{jk})\,du$. Let $\{w_1,\ldots,w_K\}$ be nonnegative weights summing to one that represent the strata weights of the target mixture population. The same weights are used for both treatment groups, so that the two standardized AHs refer to the same target population. The standardized survival function and RMST in group $j$ are
\begin{equation}
\label{eq:std_surv}
\overline{S}_j(t;\overline{\btheta}_j) = \sum_{k=1}^{K} w_k \, S_{jk}(t;\btheta_{jk}),
\qquad
\overline{R}_j(\tau;\overline{\btheta}_j) = \int_0^{\tau} \overline{S}_j(u;\overline{\btheta}_j)\,du
= \sum_{k=1}^{K} w_k \, R_{jk}(\tau;\btheta_{jk}),
\end{equation}
where $\overline{\btheta}_j=(\btheta_{j1}^\top,\ldots,\btheta_{jK}^\top)^\top$ and the standardized AH is
\begin{equation}
\label{eq:std_AH}
\overline{\eta}_j(\tau;\overline{\btheta}_j)
= \frac{1 - \overline{S}_j(\tau;\overline{\btheta}_j)}
       {\overline{R}_j(\tau;\overline{\btheta}_j)}
= \frac{\sum_{k=1}^{K} w_k \{1 - S_{jk}(\tau;\btheta_{jk})\}}
       {\sum_{k=1}^{K} w_k R_{jk}(\tau;\btheta_{jk})}.
\end{equation}
The standardized contrasts are
$$\overline{\mathrm{DAH}}(\tau)=\overline{\eta}_1(\tau;\overline{\btheta}_1)-\overline{\eta}_0(\tau;\overline{\btheta}_0)$$ and
$$\overline{\mathrm{RAH}}(\tau)=\overline{\eta}_1(\tau;\overline{\btheta}_1)/\overline{\eta}_0(\tau;\overline{\btheta}_0),$$
which are the natural estimands in a stratified analysis. Note that $\overline{\eta}_j(\tau)$ is the AH of the mixture population not the weighted average $\sum_k w_k \eta_{jk}(\tau)$ of the stratum-specific AHs,  
$\eta_{jk}(\tau)=F_{jk}(\tau;\btheta_{jk})/R_{jk}(\tau;\btheta_{jk})$. 
Let $\widehat{\btheta}_{jk}$ be the MLE of $\btheta_{jk}$, obtained by maximizing the log-likelihood \eqref{eq:loglik} based on observations in group $j$ and stratum $k$ and $\widehat{\boldsymbol{\Sigma}}_{jk}$ be the inverse of the corresponding observed information matrix. Write $\widehat{\overline{\btheta}}_j=(\widehat{\btheta}_{j1}^{\top},\ldots,\widehat{\btheta}_{jK}^{\top})^{\top}$. The plug-in estimators for AH, DAH and RAH are $$\widehat{\overline{\eta}}_j(\tau)=\overline{\eta}_j(\tau;\widehat{\overline{\btheta}}_j),~~\widehat{\overline{\mathrm{DAH}}}(\tau)=\widehat{\overline{\eta}}_1(\tau)-\widehat{\overline{\eta}}_0(\tau) ~~\mbox{and}~~\widehat{\overline{\mathrm{RAH}}}(\tau)=\widehat{\overline{\eta}}_1(\tau)/\widehat{\overline{\eta}}_0(\tau),$$ respectively. The gradient of  \eqref{eq:std_AH} with respect to $\overline{\btheta}_j$ is
\begin{equation}
\label{eq:grad_eta_std}
\nabla_{\overline{\btheta}_j}\, 
\overline{\eta}_j(\tau;\overline{\btheta}_j)=  \frac{\overline{\bD}_{j}(\tau;\overline{\btheta}_{j})}{\overline{R}_{j}(\tau; \overline{\btheta}_j)}
      + \overline{F}_{j}(\tau; \overline{\btheta}_j) \int_0^\tau \frac{\overline{\bD}_{j}(t; \overline{\btheta}_{j})dt}
       {\overline{R}_{j}(\tau; \overline{\btheta}_j)^{2}},
\end{equation}
where $F_{jk}(t; \btheta_{jk})=1-S_{jk}(t; \btheta_{jk}),$  
 $\bD_{jk}(t; \btheta_{jk})=\partial F_{jk}(t; \btheta_{jk})/\partial \btheta_{jk},$
\begin{align*}
\overline{\bD}_{j}(t;\overline{\btheta}_{j})=&\left( w_1 \bD_{j1}(t; \btheta_{j1})^{\top}, \cdots, w_K\bD_{jK}(t; \btheta_{jK})^{\top}\right)^{\top},\\
\overline{F}_{j}(t; \overline{\btheta}_j)=&\sum_{k=1}^K w_k F_{jk}(t; \btheta_{jk}) ~~~\mbox{and}~~~\overline{R}_{j}(t; \overline{\btheta}_j)= \sum_{k=1}^K w_k R_{jk}(t; \btheta_{jk}).
\end{align*}
Given the sample size $\{n_{j1}, \cdots, n_{jK}\}$, the covariance matrix of the MLE for $\overline{\btheta}_j$, $\widehat{\overline{\btheta}}_j,$ is block diagonal and can be approximated by 
$$\widehat{\overline{\bSigma}}_j=\mbox{diag}\left\{\hat{\bSigma}_{j1}, \cdots, \hat{\bSigma}_{jK}\right\}.$$
Following the delta-method, the variance of estimated AH, $\hat{\overline{\eta}}_j(\tau),$ can be estimated by 
\begin{align}
\label{eq:var_eta_std}
\widehat{\mathrm{Var}}\!\left\{\widehat{\overline{\eta}}_j(\tau)\right\}
=&\left[\nabla_{\overline{\btheta}_j}\, \overline{\eta}_j(\tau; \widehat{\overline{\btheta}}_j)\right]^{\!\top}\widehat{\overline{\bSigma}}_j\left[\nabla_{\overline{\btheta}_j}\, \overline{\eta}_j(\tau; \widehat{\overline{\btheta}}_j)\right] \nonumber \\
=& \sum_{k=1}^K w_k^2 \left[\frac{\bD_{jk}(\tau;\btheta_{jk})}{\overline{R}_{j}(\tau; \overline{\btheta}_j)}
      + \overline{F}_{j}(\tau; \overline{\btheta}_{j}) \int_0^\tau \frac{\bD_{jk}(t; \btheta_{jk})dt}
       {\overline{R}_{j}(\tau; \overline{\btheta}_j)^{2}}\right]^{\top}\widehat{\bSigma}_{jk} \nonumber \\
&\qquad\times\left[ \frac{\bD_{jk}(\tau;\btheta_{jk})}{\overline{R}_{j}(\tau; \overline{\btheta}_j)}
      + \overline{F}_{j}(\tau; \overline{\btheta}_j) \int_0^\tau \frac{\bD_{jk}(t; \btheta_{jk})dt}
       {\overline{R}_{j}(\tau; \overline{\btheta}_j)^{2}}\right]
\end{align}
The variance estimates of $\widehat{\overline{\mathrm{DAH}}}(\tau)$ and
$\log\widehat{\overline{\mathrm{RAH}}}(\tau)$ are obtained from (\ref{eq:var_eta_std}) exactly as in Section~\ref{ssec:unstrat}, and the CIs and Wald-test statistics for ${\overline{\mathrm{DAH}}}(\tau)$ and ${\overline{\mathrm{RAH}}}(\tau)$ follow accordingly. 

We treat the weights defining the target mixture population as fixed. When the target population is the empirical trial population, one may set $
w_k=(n_{1k}+n_{0k})/n,$
which is the choice used in Section~\ref{sec:example}. When the target is an external or hypothetical population with prespecified stratum proportions, the weights \(\{w_1,\ldots,w_K\}\) can be specified accordingly. If the stratum weights are instead estimated from random data, and the estimand is defined with respect to the limiting values of those estimated weights, the sampling variability of the estimated weights needs to be accounted for in the inference. We do not consider this setting in the present paper.

\subsection{Parametric families and computation}
\label{ssec:families}

The approach applies to any continuous parametric survival family for which the likelihood \eqref{eq:loglik} can be maximized and $S_j(t;\btheta_j)$, $R_j(\tau;\btheta_j)$, and the derivatives required for delta-method inference can be evaluated, analytically or numerically. We consider two families in this paper.   The Weibull family, with shape $\kappa>0$ and scale $\lambda>0$, has survival function $S(t; \kappa, \lambda)=\exp\{-(t/\lambda)^{\kappa}\}$ and a monotone hazard (constant when $\kappa=1$). For this family, $S(\tau; \kappa, \lambda)$ and its derivatives have closed forms, and $R(\tau; \kappa, \lambda)$ and its derivatives are one-dimensional integrals over $[0,\tau]$, which can be evaluated numerically (Appendix~\ref{app:weibull}). The generalized gamma family has three parameters. We use the Prentice parameterization $(\mu,\sigma,Q)$,\cite{Prentice1974} as implemented in the \textsf{R} package \texttt{flexsurv},\cite{Jackson2016} in which the Weibull family corresponds to $Q=1$ and the log-normal family to $Q=0$. The family accommodates increasing, decreasing, arc-shaped, and bathtub-shaped hazards.\cite{Cox2007} For this family, $S(\tau; \mu, \sigma, Q)$ can be evaluated via the cumulative distribution function in \texttt{flexsurv}. $R(\tau; \mu, \sigma, Q)$ can be computed by numerical integration of $S(\cdot; \mu, \sigma, Q)$ over $[0,\tau]$; and relevant derivatives can be computed by numerical differentiation (Appendix~\ref{app:gg}).

For both parametric model families, the MLE is obtained by numerical maximization of the log-likelihood function in \eqref{eq:loglik}. Appendix~\ref{app:gg} provides the required parameter transformations when the existing software uses a different parameterization. The additional shape parameter in the generalized gamma model may be difficult to estimate reliably when the number of events is small, as reflected by non-convergence, an ill-conditioned information matrix, or an extremely large standard error. We return to this issue in Sections~\ref{ssec:modelchoice} and~\ref{sec:simulation}.

\subsection{Model choice, diagnostics, and reporting}
\label{ssec:modelchoice}

Applying the proposed approach in practice requires three decisions. The first concerns the data used for model fitting: the original data or the data administratively censored at \(\tau\) (Section~\ref{ssec:unstrat}). This choice reflects the extent to which the analyst is willing to rely on the parametric model beyond \(\tau\). Using the original data with the full available follow-up is more efficient if the assumed parametric model provides an adequate fit over the entire follow-up period, whereas using administratively censored data may be preferable if the model fits the data well only over \([0,\tau]\). 

The second decision concerns the choice of parametric family. Because the model is fitted separately by treatment group and stratum, the same parametric family need not be used for every fit in the analysis. The Weibull family is nested within the generalized gamma family, so the two models can be compared using a likelihood-ratio test. More generally, goodness of fit can be assessed for each fitted model using diagnostic tools such as Cox--Snell residual plots. For more flexible parametric families, such as the generalized gamma model, it is also important to examine convergence and the estimated covariance matrix, including their stability under small perturbations of the observed data. If a simpler model, such as the Weibull model, provides an adequate fit to the observed data, it is generally preferable. Appendix~\ref{app:diag} illustrates these diagnostic procedures through examples.

The third decision concerns the choice of \(\tau\). Inference based on a parametric model can in principle be conducted at any \(\tau\), but the degree of empirical support may vary substantially. Let \(\tilde{\tau}_{jk}\) denote the largest observed time in treatment group \(j\) and stratum \(k\). When $
\tau \le \min_{j,k}\{\tilde{\tau}_{jk}\},$
\(\tau\) lies within the observed follow-up range of every group-by-stratum analysis, and no extrapolation beyond the largest observed time is required. As discussed in the Introduction, however, it may be desirable to choose a larger \(\tau\) when one or a few \(\tilde{\tau}_{jk}\) values are substantially smaller than the others. When \(\tau>\tilde{\tau}_{jk}\) for a particular group and stratum, the survival distribution over \((\tilde{\tau}_{jk},\tau]\) is entirely determined by extrapolation from the fitted parametric model. The accuracy of such extrapolation cannot be directly assessed using the observed data. We therefore recommend explicitly reporting groups and strata, together with their \(\tilde{\tau}_{jk}\) values, as illustrated in Table~\ref{table1}.

\begin{table}[!tbp]
\centering
\caption{Generalized gamma-based estimates of the average hazard with survival weight at $\tau=6,12,24,$ and $36$ months, by hormone receptor stratum and standardized, for the treatment group (N+C) and the control group (L+C), using data reconstructed from Saura et al.\cite{saura2020neratinib} AHs and their differences are in events per person-month. $^{\dagger}$At $\tau=24$ months, the L+C estimate in Stratum~A relies on model-based extrapolation, because the largest observed time in that subgroup was 13.8 months. At $\tau=36$ months, this subgroup and both N+C subgroups (largest observed times 35.9 months in Stratum~A and 30.4 months in Stratum~B) require extrapolation. The stratum-specific and standardized estimates at these truncation times therefore depend partly on the fitted distributions beyond the observed follow-up.}
\label{table1}
\footnotesize
\setlength{\tabcolsep}{6pt}
\renewcommand{\arraystretch}{1.25}
\resizebox{\textwidth}{!}{%
\begin{tabular}{l l l l l}
\toprule
Tau & & Stratum A & Stratum B & Standardized\\
\midrule
tau = 6
 & Treatment (95\%CI) & 0.099 (0.076 to 0.128) & 0.136 (0.112 to 0.165) & 0.120 (0.103 to 0.140) \\
 & Control (95\%CI) & 0.191 (0.161 to 0.226) & 0.106 (0.086 to 0.131) & 0.137 (0.120 to 0.157) \\
 & Difference (DAH) (95\%CI; p-value) & -0.092 (-0.133 to -0.051); $<$0.001 & 0.030 (-0.005 to 0.065); 0.089 & -0.017 (-0.043 to 0.009); 0.195 \\
 & Ratio (RAH) (95\%CI; p-value) & 0.519 (0.381 to 0.707); $<$0.001 & 1.283 (0.963 to 1.710); 0.089 & 0.874 (0.713 to 1.072); 0.197 \\
\midrule
tau = 12
 & Treatment (95\%CI) & 0.095 (0.076 to 0.120) & 0.119 (0.098 to 0.144) & 0.109 (0.094 to 0.126) \\
 & Control (95\%CI) & 0.205 (0.177 to 0.237) & 0.117 (0.098 to 0.140) & 0.145 (0.129 to 0.164) \\
 & Difference (DAH) (95\%CI; p-value) & -0.109 (-0.147 to -0.072); $<$0.001 & 0.002 (-0.029 to 0.033); 0.896 & -0.037 (-0.060 to -0.013); 0.002 \\
 & Ratio (RAH) (95\%CI; p-value) & 0.465 (0.355 to 0.611); $<$0.001 & 1.018 (0.784 to 1.322); 0.896 & 0.746 (0.618 to 0.902); 0.002 \\
\midrule
tau = 24$^{\dagger}$
 & Treatment (95\%CI) & 0.090 (0.072 to 0.114) & 0.109 (0.089 to 0.134) & 0.101 (0.086 to 0.118) \\
 & Control (95\%CI) & 0.209 (0.182 to 0.240) & 0.119 (0.100 to 0.141) & 0.145 (0.128 to 0.164) \\
 & Difference (DAH) (95\%CI; p-value) & -0.118 (-0.155 to -0.082); $<$0.001 & -0.009 (-0.040 to 0.021); 0.545 & -0.044 (-0.068 to -0.020); $<$0.001 \\
 & Ratio (RAH) (95\%CI; p-value) & 0.433 (0.330 to 0.569); $<$0.001 & 0.921 (0.704 to 1.205); 0.548 & 0.696 (0.570 to 0.848); $<$0.001 \\
\midrule
tau = 36$^{\dagger}$
 & Treatment (95\%CI) & 0.084 (0.065 to 0.110) & 0.102 (0.081 to 0.129) & 0.094 (0.079 to 0.112) \\
 & Control (95\%CI) & 0.209 (0.182 to 0.240) & 0.114 (0.094 to 0.137) & 0.140 (0.122 to 0.161) \\
 & Difference (DAH) (95\%CI; p-value) & -0.125 (-0.161 to -0.088); $<$0.001 & -0.012 (-0.044 to 0.020); 0.476 & -0.046 (-0.072 to -0.021); $<$0.001 \\
 & Ratio (RAH) (95\%CI; p-value) & 0.404 (0.299 to 0.544); $<$0.001 & 0.897 (0.663 to 1.213); 0.480 & 0.671 (0.536 to 0.839); $<$0.001 \\
\bottomrule
\end{tabular}%
}
\bigskip
\begin{minipage}{\textwidth}
\footnotesize
Abbreviations: AH, average hazard; DAH, difference in average hazards; RAH, ratio of average hazards; Stratum A, hormone receptor-negative disease; Stratum B, hormone receptor-positive disease; CI, confidence interval; Treatment, Neratinib + Capecitabine group; Control, Lapatinib + Capecitabine group.
All estimates were obtained with the generalized gamma model in every treatment-by-stratum subgroup.
\end{minipage}
\end{table}

None of these choices is inherently incorrect. The difficulty arises when they are made adaptively based on the observed data, which introduces a post-selection inference problem. In such cases, there is no guarantee that the inferential results presented in Section~\ref{sec:methods} remain valid when the same data are used both to select the analysis strategy and to conduct the final inference. Ideally, the parametric family and other key analysis choices should be prespecified, with sensitivity analyses conducted under reasonable alternative specifications.

In summary, when the nonparametric analysis is not adequately supported at the \(\tau\) of interest, the proposed procedure is to:
\begin{enumerate}
\item specify \(\tau\), the data used for model fitting (with or without administrative censoring at \(\tau\)), the parametric family for each treatment group and stratum, and, for stratified analyses, the stratum weights;
\item fit the model separately by treatment group and stratum;
\item compute the AH or standardized AH for each treatment group, together with the DAH and RAH, construct CIs, and conduct Wald tests of the corresponding null hypotheses (Sections~\ref{ssec:unstrat} and~\ref{ssec:strat});
\item assess goodness of fit and numerical stability for each fitted model; and
\item report the parametric family and \(\tilde{\tau}_{jk}\) for each fit, and highlight any treatment group and stratum for which model-based extrapolation is used.
\end{enumerate}

\section{Example}
\label{sec:example}

We illustrate the proposed approach using the PFS data described in Section~\ref{sec:intro}. Stratum~A (hormone receptor--negative) comprised 254 patients (126 N+C and 128 L+C), whereas Stratum~B (hormone receptor--positive) comprised 367 patients (181 N+C and 186 L+C). Although the trial used several stratification factors in randomization,\cite{saura2020neratinib} we consider only hormone receptor status as the stratification factor for illustrative purposes. Based on the reconstructed data, the estimated HRs are 0.43 (95\% CI: 0.31 to 0.58) and 1.07 (95\% CI: 0.83 to 1.38) for the hormone receptor--negative and hormone receptor--positive strata, respectively, closely agreeing with the published values of 0.42 (95\% CI: 0.31 to 0.57) and 1.08 (95\% CI: 0.84 to 1.40), respectively.\cite{saura2020neratinib} For illustration, we perform the analysis at several truncation times, \(\tau=6,12,24,\) and \(36\) months, without designating any one of them as the primary value.

Table~\ref{tab:celldiag} reports the numbers at risk and the largest observed time by treatment group and stratum. In the L+C group of Stratum~A, only four patients remained at risk at 12 months, and the largest observed time, \(\tilde{\tau}_{jk}\), was 13.8 months. Thus, inference on the AH at \(\tau=24\) and \(36\) months required model-based extrapolation (Section~\ref{ssec:modelchoice}). The largest observed time was also less than \(\tau=36\) months in the N+C groups of both strata, so inference on the AH at $\tau=36$ also required extrapolation. 
In contrast, the observed data provided insufficient support for the corresponding nonparametric analyses at these truncation times.

For the main analysis, we used the generalized gamma model for all treatment-by-stratum subgroups, with the Weibull model used in a sensitivity analysis. Each model was fitted using data administratively censored at the corresponding \(\tau\), and the results were standardized to the observed stratum distribution in the trial, with \(w_1=254/621=41\%\) and \(w_2=367/621=59\%\). 
All model fits to the observed data converged.
The condition numbers of the estimated covariance matrices ranged from 18 to 145,
and the proportion of bootstrap resamples in which the generalized gamma model could not be refitted did not exceed 1.5\% in any subgroup 
(Table~\ref{tab:celldiag}). In addition, likelihood-ratio tests of \(Q=1\) favored the generalized gamma model over the Weibull model in one treatment-by-stratum subgroup at \(\tau=6\) months, three subgroups at \(\tau=12\) months, and all four subgroups at \(\tau=24\) and \(36\) months. The Cox--Snell residual plots also suggested a better fit of the generalized gamma model than the Weibull model, particularly in the two N+C subgroups (Figure~\ref{fig:coxsnell}).

Table~\ref{table1} presents the results based on the generalized gamma model; DAH \(<0\) and RAH \(<1\) favor N+C. In Stratum~A, the AH for N+C was lower than that for L+C at all truncation times, with the RAH ranging from 0.52 at \(\tau=6\) months to 0.40 at \(\tau=36\) months. In Stratum~B, the AH for N+C was higher than that for L+C at \(\tau=6\) months (RAH 1.28), whereas the AHs in the two groups were similar at the other truncation times (RAHs 1.02, 0.92, and 0.90 at \(\tau=12\), 24, and 36 months, respectively), consistent with the non-PH pattern shown in Figure~\ref{fig:KM}. At \(\tau=6\) months, the AH for N+C was only slightly lower than that for L+C because of the contrasting treatment effects across the two strata (RAH 0.874, 95\% CI: 0.713 to 1.072). At \(\tau=12\) months, N+C had a significantly lower AH than L+C (DAH \(-0.037\), 95\% CI: \(-0.060\) to \(-0.013\); RAH 0.746, 95\% CI: 0.618 to 0.902). The AH estimates at \(\tau=24\) and \(36\) months rely on model-based extrapolation in this analysis.

Under the Weibull model (Table~\ref{tab:wbsens}), the RAH was 0.878 at \(\tau=6\) months and 0.753 at \(\tau=12\) months, nearly identical to the corresponding estimates from the generalized gamma model. At \(\tau=24\) and \(36\) months, however, the RAHs were 0.723 and 0.728, respectively, which differed noticeably from those obtained under the generalized gamma model. These results suggest that analyses requiring little or no extrapolation are more robust to the choice of parametric model than analyses that rely more heavily on extrapolation.

Based on the estimated AH at \(\tau=12\) months, there were 14.5 and 10.9 PFS events per 100 progression-free person-months in the L+C and N+C groups, respectively, corresponding to an absolute reduction of approximately 3.7 events per 100 progression-free person-months. Equivalently, the reciprocals of the AH estimates indicate that approximately 6.9 and 9.2 progression-free person-months were accumulated per expected PFS event in the L+C and N+C groups, respectively. Because the data were reconstructed from published Kaplan--Meier curves and the analysis was conducted post hoc, these results are intended solely to illustrate the proposed approach.

\section{Numerical Studies}
\label{sec:simulation}
\subsection{Design}
\label{ssec:sim-config-supp}

The numerical studies had two aims: to illustrate how nonparametric inference deteriorates as the risk set at \(\tau\) becomes small or is completely exhausted, and to evaluate the finite-sample bias and coverage of the proposed parametric procedure in such settings. All event times were generated from Weibull models. Thus, both the Weibull and generalized gamma models were correctly specified, although the generalized gamma model included one unnecessary shape parameter.

The simulation design mimics the real-data example in Section~\ref{sec:example}. Event times in the four treatment-by-stratum subgroups were generated from Weibull distributions, with shape and scale parameters set to be the MLEs from the reconstructed data: \((1.12,11.18)\) for N+C and \((1.65,5.27)\) for L+C in Stratum~A, and \((1.03,8.76)\) for N+C and \((1.28,9.04)\) for L+C in Stratum~B. Censoring times were either generated from a Weibull distribution with \((\kappa,\lambda)=(1.11,17.79)\), corresponding to the MLEs from the reconstructed censoring distribution, or no censoring was imposed. Two total sample sizes, \(n=600\) and \(1{,}000\), were considered. 
Under the first allocation rule, the number of patients in Stratum~A was drawn from a binomial distribution with probability 0.4, and 40\% of the patients in Stratum~A and 60\% of those in Stratum~B were assigned to N+C. The stratum-specific weights were fixed at 0.4 and 0.6 for Strata~A and B, respectively. Combining the two sample sizes, two censoring settings, and four truncation times, \(\tau=6,12,24,\) and \(36\) months, yielded 16 simulation configurations. The simulation results under the second allocation rule, in which the number of patients in Stratum~A was drawn from a binomial distribution with probability 0.7 and the two treatment groups were allocated 1:1 within each stratum, are reported in Appendix~\ref{app:sim}.

In each of 3,000 replications, we conducted both the nonparametric and parametric stratified analyses. For the AH in each treatment group, the DAH, and \(\log(\mathrm{RAH})\), we calculated the bias and empirical standard error of their estimators, as well as the empirical coverage probability of the corresponding 95\% CIs for the true values defined by \eqref{eq:std_AH}. Replications in which the risk set was depleted at \(\tau\) in at least one treatment-by-stratum subgroup were flagged. A nonparametric point estimate may still be obtainable in some such replications, but the observed data do not support nonparametric inference on AH over the full interval \([0,\tau]\). Flagged replications were excluded from summaries of the performance of the nonparametric procedure, but their numbers were reported. 
A parametric fit that returned an error (non-convergence) in any subgroup was counted as a failure of that procedure in that replication.
If more than 2\% of replications were flagged or failed, we did not summarize the performance of the corresponding procedure for that configuration, because the remaining replications would constitute a selected subset. The 2\% threshold was used as a reporting convention. 
To characterize the sparsity of the risk set before it is exhausted, we also report \(\overline{m}_j(\tau)\).
For each replication, $m_j(\tau)$ wad defined as the smallest risk-set size across strata in group $j$ up to $\tau.$
We calculated \(\overline{m}_j(\tau)\) by averaging $m_j(\tau)$ over all 3,000 replications, including flagged one.

\subsection{Results}

Tables~\ref{tab:sim-p3-cens1} and~\ref{tab:sim-p3-cens0} report the results under random censoring and no censoring, respectively. With random censoring and \(n=600\), \(\overline{m}_j(\tau)\) in the control group decreased from 32.0 at \(\tau=6\) months to 2.5 at \(\tau=12\) months and 1.0 thereafter.  657, 2,999, and 3,000 of the 3,000 replications were flagged at \(\tau=12\), 24, and 36 months, respectively; the corresponding numbers for \(n=1{,}000\) were 236, 3,000, and 3,000. Without censoring, the risk sets were larger, but 171, 2,997, and 3,000 replications for \(n=600\), and 32, 2,994, and 3,000 replications for \(n=1{,}000\) were still flagged at $\tau=12$, 24 and 36 months. Thus, increasing the sample size or eliminating censoring delayed risk-set exhaustion but did not prevent it. The Weibull model converged in every replication, whereas the generalized gamma model failed to converge in at most 37 of the 3,000 replications.

\begin{table}[!tbp]
\centering
\caption{Bias and coverage of the nonparametric and parametric procedures for the standardized AH, the DAH, and $\log(\mathrm{RAH})$, based on 3,000 replications under censoring pattern~II (no censoring). $\overline{m}_j(\tau)$ and the flag are defined in Section~\ref{ssec:sim-config-supp}. Nonparametric summaries are conditional on the absence of the flag; generalized gamma summaries are conditional on convergence.}
\label{tab:sim-p3-cens0}
\footnotesize
\setlength{\tabcolsep}{3pt}
\renewcommand{\arraystretch}{1.16}
\resizebox{\textwidth}{!}{%
\begin{tabular}{p{1.8cm}p{1.8cm}cccccccc}
\toprule
\multicolumn{2}{c}{sample size} & \multicolumn{4}{c}{N = 600} & \multicolumn{4}{c}{N = 1000} \\
\cmidrule(lr){3-6}\cmidrule(lr){7-10}
\multicolumn{2}{c}{tau} & 6 & 12 & 24 & 36 & 6 & 12 & 24 & 36 \\
\midrule
\multicolumn{10}{l}{\textbf{Average minimum positive risk-set size up to $\tau$, $\overline{m}_j(\tau)$}} \\
AH treatment & & 59.28 & 33.37 & 9.47 & 2.7 & 97.88 & 54.87 & 15.54 & 4.16 \\
AH control & & 42.64 & 3.85 & 1 & 1 & 70.47 & 5.8 & 1 & 1 \\
\addlinespace[2pt]
\multicolumn{10}{l}{\textbf{True value}} \\
AH treatment & & 0.099 & 0.102 & 0.105 & 0.105 & 0.099 & 0.102 & 0.105 & 0.105 \\
AH control & & 0.125 & 0.14 & 0.144 & 0.145 & 0.125 & 0.14 & 0.144 & 0.145 \\
DAH & & -0.026 & -0.038 & -0.039 & -0.039 & -0.026 & -0.038 & -0.039 & -0.039 \\
logRAH & & -0.229 & -0.314 & -0.319 & -0.317 & -0.229 & -0.314 & -0.319 & -0.317 \\
\addlinespace[2pt]
\rowcolor[gray]{0.9}
\multicolumn{2}{>{\cellcolor[gray]{0.9}}p{3.8cm}}{No. of replications with an empty risk set\newline at $\tau$ in some cell} & 0 & 171 & 2997 & 3000 & 0 & 32 & 2994 & 3000 \\
\multicolumn{2}{p{3.6cm}}{\# replications where the Weibull fit did not converge} & 0 & 0 & 0 & 0 & 0 & 0 & 0 & 0 \\
\multicolumn{2}{p{3.6cm}}{\# replications where the generalized gamma fit did not converge} & 0 & 0 & 0 & 1 & 0 & 0 & 0 & 0 \\
\midrule
\multicolumn{10}{l}{\textbf{Bias (MCSE) [$\times 10^{-3}$]}} \\
\multirow[t]{3}{*}{logRAH} & nonparametric & -0.57 (2.03) & - & - & - & 2.45 (1.59) & 0.96 (1.25) & - & - \\
& Weibull & -0.64 (2.02) & 0.09 (1.62) & -0.04 (1.39) & -0.07 (1.34) & 2.51 (1.59) & 1.13 (1.24) & 1.18 (1.08) & 0.92 (1.06) \\
& GG & -0.33 (2.02) & 1.56 (1.61) & -0.01 (1.39) & 0.02 (1.34) & 3.01 (1.62) & 1.01 (1.24) & 1.21 (1.08) & 0.99 (1.06) \\
\multirow[t]{3}{*}{DAH} & nonparametric & -0.02 (0.22) & - & - & - & 0.29 (0.18) & 0.09 (0.15) & - & - \\
& Weibull & -0.03 (0.22) & -0.01 (0.19) & -0.08 (0.17) & -0.10 (0.17) & 0.30 (0.18) & 0.11 (0.15) & 0.09 (0.13) & 0.05 (0.13) \\
& GG & -0.00 (0.22) & 0.19 (0.19) & -0.07 (0.17) & -0.07 (0.17) & 0.33 (0.18) & 0.10 (0.15) & 0.09 (0.13) & 0.06 (0.13) \\
\multirow[t]{3}{*}{AH treatment} & nonparametric & 0.25 (0.15) & - & - & - & 0.30 (0.12) & 0.26 (0.10) & - & - \\
& Weibull & 0.24 (0.15) & 0.35 (0.13) & 0.34 (0.11) & 0.35 (0.11) & 0.29 (0.12) & 0.27 (0.10) & 0.29 (0.08) & 0.29 (0.08) \\
& GG & 0.28 (0.15) & 0.38 (0.13) & 0.32 (0.11) & 0.33 (0.11) & 0.34 (0.12) & 0.25 (0.10) & 0.29 (0.08) & 0.28 (0.08) \\
\multirow[t]{3}{*}{AH control} & nonparametric & 0.27 (0.16) & - & - & - & 0.01 (0.13) & 0.17 (0.11) & - & - \\
& Weibull & 0.26 (0.16) & 0.36 (0.14) & 0.42 (0.13) & 0.45 (0.13) & -0.01 (0.13) & 0.16 (0.11) & 0.21 (0.10) & 0.24 (0.10) \\
& GG & 0.28 (0.16) & 0.19 (0.14) & 0.39 (0.13) & 0.40 (0.13) & 0.01 (0.13) & 0.16 (0.11) & 0.20 (0.10) & 0.22 (0.10) \\
\midrule
\multicolumn{10}{l}{\textbf{Coverage}} \\
\multirow[t]{3}{*}{logRAH} & nonparametric & 0.963 & - & - & - & 0.953 & 0.993 & - & - \\
& Weibull & 0.957 & 0.950 & 0.952 & 0.949 & 0.949 & 0.953 & 0.955 & 0.954 \\
& GG & 0.959 & 0.952 & 0.953 & 0.948 & 0.946 & 0.953 & 0.957 & 0.954 \\
\multirow[t]{3}{*}{DAH} & nonparametric & 0.964 & - & - & - & 0.954 & 0.995 & - & - \\
& Weibull & 0.957 & 0.951 & 0.954 & 0.951 & 0.946 & 0.952 & 0.954 & 0.953 \\
& GG & 0.957 & 0.949 & 0.955 & 0.949 & 0.946 & 0.953 & 0.954 & 0.951 \\
\multirow[t]{3}{*}{AH treatment} & nonparametric & 0.952 & - & - & - & 0.953 & 0.952 & - & - \\
& Weibull & 0.951 & 0.944 & 0.949 & 0.950 & 0.953 & 0.951 & 0.951 & 0.953 \\
& GG & 0.952 & 0.943 & 0.950 & 0.949 & 0.953 & 0.953 & 0.950 & 0.952 \\
\multirow[t]{3}{*}{AH control} & nonparametric & 0.966 & - & - & - & 0.962 & 0.999 & - & - \\
& Weibull & 0.955 & 0.950 & 0.948 & 0.950 & 0.952 & 0.946 & 0.955 & 0.953 \\
& GG & 0.954 & 0.950 & 0.948 & 0.948 & 0.951 & 0.947 & 0.954 & 0.953 \\
\bottomrule
\end{tabular}}
\end{table}

\begin{table}[!tbp]
\centering
\caption{Bias and coverage of the nonparametric and parametric procedures for the standardized AH, the DAH, and $\log(\mathrm{RAH})$, based on 3,000 replications under censoring pattern~I (random censoring). $\overline{m}_j(\tau)$ and the flag are defined in Section~\ref{ssec:sim-config-supp}. Nonparametric summaries are conditional on the absence of the flag; generalized gamma summaries are conditional on convergence.}
\label{tab:sim-p3-cens1}
\footnotesize
\setlength{\tabcolsep}{3pt}
\renewcommand{\arraystretch}{1.16}
\resizebox{\textwidth}{!}{%
\begin{tabular}{p{1.8cm}p{1.8cm}cccccccc}
\toprule
\multicolumn{2}{c}{sample size} & \multicolumn{4}{c}{N = 600} & \multicolumn{4}{c}{N = 1000} \\
\cmidrule(lr){3-6}\cmidrule(lr){7-10}
\multicolumn{2}{c}{tau} & 6 & 12 & 24 & 36 & 6 & 12 & 24 & 36 \\
\midrule
\multicolumn{10}{l}{\textbf{Average minimum positive risk-set size up to $\tau$, $\overline{m}_j(\tau)$}} \\
AH treatment & & 44.3 & 17.98 & 2.71 & 1.06 & 73.03 & 29.33 & 4.14 & 1.16 \\
AH control & & 32.04 & 2.54 & 1 & 1 & 52.53 & 3.53 & 1 & 1 \\
\addlinespace[2pt]
\multicolumn{10}{l}{\textbf{True value}} \\
AH treatment & & 0.099 & 0.102 & 0.105 & 0.105 & 0.099 & 0.102 & 0.105 & 0.105 \\
AH control & & 0.125 & 0.14 & 0.144 & 0.145 & 0.125 & 0.14 & 0.144 & 0.145 \\
DAH & & -0.026 & -0.038 & -0.039 & -0.039 & -0.026 & -0.038 & -0.039 & -0.039 \\
logRAH & & -0.229 & -0.314 & -0.319 & -0.317 & -0.229 & -0.314 & -0.319 & -0.317 \\
\addlinespace[2pt]
\rowcolor[gray]{0.9}
\multicolumn{2}{>{\cellcolor[gray]{0.9}}p{3.8cm}}{No. of replications with an empty risk set\newline at $\tau$ in some cell} & 0 & 657 & 2999 & 3000 & 0 & 236 & 3000 & 3000 \\
\multicolumn{2}{p{3.6cm}}{\# replications where the Weibull fit did not converge} & 0 & 0 & 0 & 0 & 0 & 0 & 0 & 0 \\
\multicolumn{2}{p{3.6cm}}{\# replications where the generalized gamma fit did not converge} & 0 & 0 & 23 & 37 & 1 & 0 & 2 & 6 \\
\midrule
\multicolumn{10}{l}{\textbf{Bias (MCSE) [$\times 10^{-3}$]}} \\
\multirow[t]{3}{*}{logRAH} & nonparametric & 2.42 (2.25) & - & - & - & -0.81 (1.70) & - & - & - \\
& Weibull & 2.22 (2.24) & 1.39 (1.85) & 1.22 (1.76) & 0.87 (1.74) & -0.71 (1.69) & 0.31 (1.41) & 0.28 (1.34) & -0.25 (1.33) \\
& GG & 2.40 (2.25) & 1.75 (1.86) & 1.34 (1.80) & 0.95 (1.81) & -0.59 (1.69) & 0.46 (1.41) & 0.50 (1.37) & -0.20 (1.38) \\
\multirow[t]{3}{*}{DAH} & nonparametric & 0.30 (0.25) & - & - & - & -0.07 (0.19) & - & - & - \\
& Weibull & 0.27 (0.25) & 0.14 (0.22) & 0.08 (0.22) & 0.07 (0.22) & -0.07 (0.19) & 0.01 (0.17) & -0.02 (0.17) & -0.07 (0.16) \\
& GG & 0.29 (0.25) & 0.20 (0.22) & 0.12 (0.22) & 0.15 (0.22) & -0.06 (0.19) & 0.04 (0.17) & 0.02 (0.17) & -0.04 (0.17) \\
\multirow[t]{3}{*}{AH treatment} & nonparametric & 0.40 (0.16) & - & - & - & 0.12 (0.13) & - & - & - \\
& Weibull & 0.44 (0.16) & 0.41 (0.14) & 0.44 (0.14) & 0.35 (0.14) & 0.15 (0.13) & 0.25 (0.11) & 0.30 (0.11) & 0.26 (0.11) \\
& GG & 0.44 (0.16) & 0.39 (0.14) & 0.39 (0.14) & 0.22 (0.14) & 0.15 (0.13) & 0.24 (0.11) & 0.30 (0.11) & 0.22 (0.11) \\
\multirow[t]{3}{*}{AH control} & nonparametric & 0.10 (0.18) & - & - & - & 0.20 (0.14) & - & - & - \\
& Weibull & 0.17 (0.18) & 0.27 (0.17) & 0.36 (0.17) & 0.28 (0.17) & 0.22 (0.14) & 0.24 (0.13) & 0.33 (0.13) & 0.33 (0.13) \\
& GG & 0.15 (0.18) & 0.19 (0.17) & 0.27 (0.17) & 0.07 (0.17) & 0.21 (0.14) & 0.20 (0.13) & 0.29 (0.13) & 0.26 (0.13) \\
\midrule
\multicolumn{10}{l}{\textbf{Coverage}} \\
\multirow[t]{3}{*}{logRAH} & nonparametric & 0.952 & - & - & - & 0.956 & - & - & - \\
& Weibull & 0.945 & 0.943 & 0.943 & 0.943 & 0.950 & 0.955 & 0.947 & 0.947 \\
& GG & 0.947 & 0.946 & 0.943 & 0.948 & 0.952 & 0.954 & 0.949 & 0.948 \\
\multirow[t]{3}{*}{DAH} & nonparametric & 0.953 & - & - & - & 0.960 & - & - & - \\
& Weibull & 0.944 & 0.940 & 0.942 & 0.946 & 0.954 & 0.953 & 0.949 & 0.946 \\
& GG & 0.945 & 0.939 & 0.941 & 0.945 & 0.952 & 0.952 & 0.946 & 0.943 \\
\multirow[t]{3}{*}{AH treatment} & nonparametric & 0.950 & - & - & - & 0.947 & - & - & - \\
& Weibull & 0.953 & 0.948 & 0.948 & 0.951 & 0.951 & 0.948 & 0.948 & 0.948 \\
& GG & 0.954 & 0.951 & 0.948 & 0.952 & 0.950 & 0.949 & 0.947 & 0.941 \\
\multirow[t]{3}{*}{AH control} & nonparametric & 0.957 & - & - & - & 0.959 & - & - & - \\
& Weibull & 0.948 & 0.944 & 0.936 & 0.937 & 0.945 & 0.944 & 0.946 & 0.949 \\
& GG & 0.945 & 0.942 & 0.936 & 0.939 & 0.947 & 0.945 & 0.943 & 0.945 \\
\bottomrule
\end{tabular}}
\end{table}

Both parametric estimators showed little bias, and the empirical coverage probabilities of the constructed CIs were generally close to the nominal 95\% level. Under random censoring, the coverage probabilities ranged from 0.936 to 0.955 for the Weibull model and from 0.936 to 0.954 for the generalized gamma model; without censoring, the corresponding ranges were 0.944 to 0.957 and 0.943 to 0.959. At \(\tau=6\) months, where the nonparametric procedure was never flagged, its bias and coverage were similar to those of the parametric procedures. In contrast, with no censoring, \(n=1{,}000\), and \(\tau=12\) months, 32 of the 3,000 replications were flagged. Among remaining replications, the conditional coverage probabilities of the nonparametric procedure were 0.999 for the AH in the control group, 0.995 for the DAH, and 0.993 for the RAH, compared with approximately 0.95 for both parametric models.

In summary, under correct model specification, CIs from the parametric procedures maintained coverage close to the nominal level. In contrast, the nonparametric procedure became increasingly conservative as the risk sets became smaller, and an increasing proportion of replications were flagged because of depleted risk sets. The generalized gamma model performed similarly to the Weibull model, although with occasional convergence failures. The alternative stratum allocation ratio of 0.7 versus 0.3 yielded similar findings (Appendix~\ref{app:sim}).

\section{Discussion}
\label{sec:discussion}
The contribution of this paper is practical rather than methodological. The nonparametric procedure should remain the default when it is adequately supported by the observed data. However, when nonparametric inference is not supported by observed data or even infeasible at the truncation time of clinical interest, the analyst may otherwise be forced to choose an earlier \(\tau\), potentially limiting the clinical relevance of the estimand and discarding substantial follow-up information available in other treatment-by-stratum subgroups. This represents an important practical limitation of the nonparametric inference, particularly in stratified analyses. We therefore propose a parametric alternative for such settings. 

The proposed approach parallels the use of parametric models for RMST,\cite{Royston:2011ci,Royston2002-kl} but targets different estimands. The AH is defined on the event-rate scale, whereas the RMST is defined on the time scale, and the RAH is more directly comparable in interpretation to a hazard ratio. The risk-set requirement is also more restrictive for the AH. Under mild conditions, the RMST can be estimated nonparametrically for \(\tau\) as large as the largest observed follow-up time,\cite{Tian2020-oz} whereas the inference on AH additionally requires  the estimation of survival probability at \(\tau\) be adequately supported by the observed follow-up.

The proposed procedure requires specification of a parametric survival model within each treatment group and stratum, which may be restrictive in practice. The goodness of fit of the assumed model can be assessed empirically using the observed data (Section~\ref{ssec:modelchoice}). However, when inference on the AH relies on model-based extrapolation beyond the observed follow-up in some subgroups, as may occur in the settings for which the proposed method is intended, model adequacy cannot be evaluated over the entire time window of interest. Our numerical studies showed that inference based on such extrapolation can be sensitive to the assumed parametric model. Therefore, whenever extrapolation is required, it should be acknowledged explicitly and the resulting estimates and inference should be interpreted with appropriate caution.

The proposed approach can be extended in several directions. First, different parametric families may be used for different treatment-by-stratum subgroups. When the data provide adequate support, a nonparametric estimator may also be used in selected subgroups, potentially improving the robustness of the overall analysis. Second, when extrapolation is required in a particular subgroup, the survival distribution may be estimated using a hybrid nonparametric--parametric approach. For example, the AH estimator may be based on the survival function
\[
\widehat{S}(t)
=
\widehat{S}_{KM}(t)I(t<b)
+
\widehat{S}_{KM}(b)
\frac{S(t;\widehat{\btheta})}
     {S(b;\widehat{\btheta})}
I(t\geq b),
\]
where \(\widehat{S}_{KM}(\cdot)\) and \(S(\cdot;\widehat{\btheta})\) denote the Kaplan--Meier and parametric estimators of the survival function, respectively, and \(b\) is a time point up to which the Kaplan--Meier estimator is adequately supported by the observed follow-up. This hybrid approach reduces reliance on the parametric model over the observed portion of the survival curve and makes the contribution of model-based extrapolation more explicit. For both extensions, corresponding statistical inference procedures can be developed as relatively straightforward extensions of those proposed in the current paper.

Several limitations should be noted. First, the numerical studies evaluated finite-sample performance under correct model specification rather than robustness to model misspecification. The latter is beyond the scope of the present study and warrants further investigation, including the use of more flexible parametric models\cite{Royston2002-kl} and the extensions described above. Second, the numerical studies used a common parametric family across all treatment-by-stratum subgroups. We did not evaluate the performance of inference following more flexible, data-adaptive model selection or model building within individual subgroups. Third, the proposed inference is asymptotic within each treatment-by-stratum subgroup and therefore requires adequate information in every subgroup. In applications with sparse strata, collapsing strata may be considered, but we did not investigate such strategies in this study.

The proposed parametric methods, including those based on the Weibull and generalized gamma distributions, are implemented in the \textsf{R} package \texttt{survAHparam} (version 0.2.0), which is available upon request to the authors, whereas the nonparametric methods are implemented in the \textsf{R} package \texttt{survAH}.\cite{uno2025survah} These packages allow different models to be used conveniently across treatment-by-stratum subgroups, as discussed above.



\subsection*{Funding}
Research reported in this publication was supported by the National Institute of General Medical Sciences of the National Institutes of Health under award number R01GM152499 and the National Heart Lung and Blood Institute
 of the National Institutes of Health under award number R01HL089778.


\appendix
\section*{Appendix}
\renewcommand{\thetable}{\thesection\arabic{table}}
\renewcommand{\thefigure}{\thesection\arabic{figure}}

\section{Computational details\label{app:parm-details}}
\setcounter{table}{0}\setcounter{figure}{0}

This appendix gives the expressions used to compute $S(\tau)$, $R(\tau)$, and their derivatives for the two families, as implemented in the \textsf{R} package \texttt{survAHparam}.

\subsection{Weibull family\label{app:weibull}}

For shape $\kappa$ and scale $\lambda$, let $c=(\tau/\lambda)^{\kappa}$. Then $S(\tau)=e^{-c}$ and
\[
\frac{\partial S(\tau)}{\partial\lambda}=\frac{\kappa}{\lambda}\,c\,e^{-c},
\qquad
\frac{\partial S(\tau)}{\partial\kappa}=-\,c\,\log\!\left(\frac{\tau}{\lambda}\right)e^{-c}.
\]
The RMST and its derivatives, obtained by differentiating under the integral sign, are
\begin{align*}
R(\tau)&=\int_0^\tau \exp\{-(u/\lambda)^{\kappa}\}\,du,\\
\frac{\partial R(\tau)}{\partial\lambda}&=\frac{\kappa}{\lambda^{\kappa+1}}\int_0^\tau u^{\kappa}\exp\{-(u/\lambda)^{\kappa}\}\,du,\\
\frac{\partial R(\tau)}{\partial\kappa}&=\int_0^\tau \left(\frac{u}{\lambda}\right)^{\kappa}\log\!\left(\frac{\lambda}{u}\right)\exp\{-(u/\lambda)^{\kappa}\}\,du .
\end{align*}
These one-dimensional integrals can be written in terms of the incomplete gamma function and its derivative with respect to the shape argument, but we evaluate them directly by adaptive numerical quadrature over $[0,\tau]$ (\texttt{stats::integrate}). The MLE is obtained by numerical maximization of \eqref{eq:loglik} over $(\log\kappa,\log\lambda)$, and the covariance matrix of $(\widehat{\kappa},\widehat{\lambda})$ is the inverse of the observed information matrix on the $(\kappa,\lambda)$ scale, whose elements are the closed-form second derivatives of the Weibull log-likelihood. The gradient in \eqref{eq:grad_eta} is evaluated on the same scale.

\subsection{Generalized gamma family\label{app:gg}}

We use the Prentice parameterization $(\mu,\sigma,Q)$ of \texttt{flexsurv}. The survival function is evaluated as $S(t)=1-\texttt{pgengamma}(t;\mu,\sigma,Q)$, which is valid for all values of $Q$; several fitted subgroups in the example have $\widehat{Q}<0$ (Table~\ref{tab:celldiag}). $R(\tau)$ is obtained by adaptive numerical integration of $S$ over $[0,\tau]$ (\texttt{stats::integrate}, relative tolerance $10^{-8}$). The gradient of $\eta(\tau)$ or, in a stratified analysis, of $\overline{\eta}_j(\tau)$ with respect to the stacked parameter vector is computed by numerical differentiation with Richardson extrapolation (\texttt{numDeriv::grad}\cite{numDeriv}); the latter is numerically equivalent to \eqref{eq:grad_eta_std}. Because \texttt{flexsurv} maximizes the likelihood on the scale $(Q,\log\sigma,\mu)$, its covariance matrix is transformed to the scale $(Q,\sigma,\mu)$, on which the gradient is taken, by the delta method with Jacobian $\mathrm{diag}(1,\sigma,1)$.
\section{Diagnostics for the example\label{app:diag}}
\setcounter{table}{0}\setcounter{figure}{0}

\begin{table}[!tbp]
\centering
\caption{Diagnostics for the fits in each treatment-by-stratum subgroup of the example. Events are counted after administrative censoring at $\tau$, and the number at risk just before $\tau$. LR, likelihood-ratio statistic for $H_0\colon Q=1$ (1 df); Cond., condition number of the estimated generalized gamma covariance matrix; Boot.\ fail, percentage of 200 bootstrap resamples in which the generalized gamma model could not be refitted. The largest observed times $\tilde{\tau}_{jk}$ were 13.8 (L+C, Stratum~A), 35.9 (N+C, Stratum~A), 36.0 (L+C, Stratum~B), and 30.4 (N+C, Stratum~B) months.}
\label{tab:celldiag}
\footnotesize
\setlength{\tabcolsep}{4pt}
\begin{tabular}{clcccrrrrcrc}
\toprule
& & & & At & \multicolumn{2}{c}{Log-likelihood} & & & & & Boot. \\
\cmidrule(lr){6-7}
$\tau$ & Stratum & Arm & Events & risk & Weibull & GG & LR & $p$ & $\widehat{Q}$ (95\% CI) & Cond. & fail (\%) \\
\midrule
6 & A & L+C & 90 & 24 & -220.4 & -219.7 & 1.46 & 0.227 & 0.35 ($-0.58$, 1.27) & 72 & 0.0 \\
 & A & N+C & 50 & 52 & -160.4 & -159.0 & 2.95 & 0.086 & $-0.17$ ($-1.49$, 1.14) & 100 & 1.5 \\
 & B & L+C & 73 & 58 & -231.3 & -231.2 & 0.10 & 0.755 & 0.77 ($-0.46$, 1.99) & 145 & 0.0 \\
 & B & N+C & 94 & 61 & -276.7 & -271.1 & 11.12 & $<$0.001 & $-0.42$ ($-1.19$, 0.36) & 51 & 0.5 \\
\addlinespace[2pt]
12 & A & L+C & 106 & 4 & -260.4 & -256.6 & 7.59 & 0.006 & 0.15 ($-0.46$, 0.76) & 30 & 0.0 \\
 & A & N+C & 70 & 26 & -232.8 & -227.5 & 10.64 & 0.001 & $-0.51$ ($-1.38$, 0.36) & 46 & 0.5 \\
 & B & L+C & 103 & 20 & -321.7 & -320.2 & 3.04 & 0.081 & 0.45 ($-0.13$, 1.03) & 29 & 0.0 \\
 & B & N+C & 113 & 28 & -350.8 & -338.6 & 24.32 & $<$0.001 & $-0.67$ ($-1.26$, $-0.08$) & 32 & 0.0 \\
\addlinespace[2pt]
24 & A & L+C & 108 & 0 & -263.7 & -260.2 & 6.89 & 0.009 & 0.22 ($-0.36$, 0.81) & 28 & 0.0 \\
 & A & N+C & 81 & 4 & -272.9 & -266.0 & 13.83 & $<$0.001 & $-0.39$ ($-1.10$, 0.32) & 32 & 0.5 \\
 & B & L+C & 114 & 2 & -357.0 & -354.1 & 5.72 & 0.017 & 0.40 ($-0.08$, 0.88) & 20 & 0.0 \\
 & B & N+C & 127 & 5 & -401.2 & -387.6 & 27.21 & $<$0.001 & $-0.52$ ($-1.04$, 0.01) & 27 & 0.0 \\
\addlinespace[2pt]
36 & A & L+C & 108 & 0 & -263.7 & -260.2 & 6.89 & 0.009 & 0.22 ($-0.36$, 0.81) & 28 & 0.0 \\
 & A & N+C & 82 & 0 & -279.0 & -270.6 & 16.80 & $<$0.001 & $-0.44$ ($-1.11$, 0.24) & 29 & 0.5 \\
 & B & L+C & 114 & 1 & -361.6 & -356.5 & 10.26 & 0.001 & 0.27 ($-0.18$, 0.71) & 18 & 0.0 \\
 & B & N+C & 128 & 0 & -405.4 & -391.4 & 28.04 & $<$0.001 & $-0.51$ ($-1.02$, 0.01) & 26 & 0.0 \\
\bottomrule
\end{tabular}
\end{table}

\begin{figure}[!tbp]
\centering
\includegraphics[width=\linewidth]{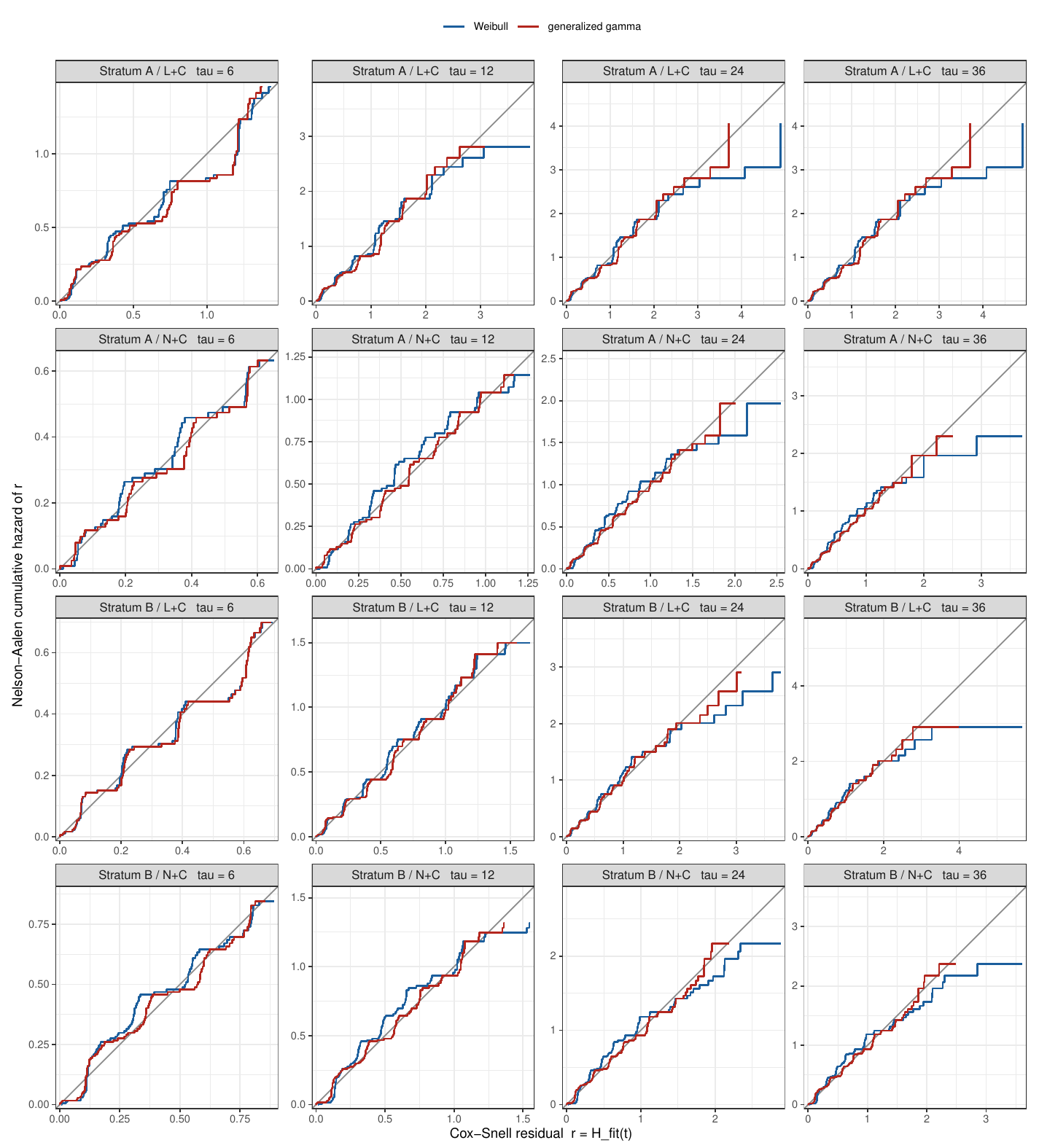}
\caption{Cox--Snell residual plots for the example, by treatment-by-stratum subgroup (rows) and truncation time $\tau$ (columns), for the Weibull and generalized gamma fits to the data censored at $\tau$. A well-fitting model follows the $45^\circ$ line (grey). N+C, neratinib plus capecitabine; L+C, lapatinib plus capecitabine; Stratum~A, hormone receptor--negative; Stratum~B, hormone receptor--positive.}
\label{fig:coxsnell}
\end{figure}

Table~\ref{tab:celldiag} reports diagnostics for the fits used in Section~\ref{sec:example}; each treatment-by-stratum subgroup was fitted to the data administratively censored at $\tau$. LR is the likelihood-ratio statistic for $Q=1$ within the subgroup (one degree of freedom). Cond.\ is the ratio of the largest to the smallest eigenvalue of the estimated covariance matrix of the generalized gamma parameters on the scale used for estimation in \texttt{flexsurv}, $(Q,\log\sigma,\mu)$; it depends on the scaling of the parameters and should be read as a coarse indicator. Boot.\ fail is the percentage of 200 nonparametric bootstrap resamples of the subgroup in which the generalized gamma model could not be refitted, that is, the resample contained fewer than three events or the refit did not return a finite log-likelihood, finite estimates, and a positive-definite covariance matrix; with 200 resamples, it is resolved only to 0.5\%. Figure~\ref{fig:coxsnell} shows Cox--Snell residual plots for the same fits: if the fitted model is adequate, the Nelson--Aalen estimate of the cumulative hazard of the residuals $r_i=-\log\widehat{S}(X_i),$ where $X_i$ is the observed time of subject $i,$ follows the $45^\circ$ line. Table~\ref{tab:wbsens} gives the standardized estimates with the Weibull model in every subgroup, for comparison with Table~\ref{table1}.

\begin{table}[!tbp]
\centering
\caption{Standardized AH estimates for the example with the Weibull model in every cell, compared with the generalized gamma estimates of Table~\ref{table1}. AHs are in events per person-month.}
\label{tab:wbsens}
\footnotesize
\setlength{\tabcolsep}{4pt}
\begin{tabular}{clcccc}
\toprule
$\tau$ & Family & Std.\ AH (N+C) & Std.\ AH (L+C) & DAH ($p$) & RAH ($p$) \\
\midrule
6  & Weibull          & 0.121 & 0.138 & $-0.017$ ($0.198$) & 0.878 ($0.201$) \\
   & generalized gamma & 0.120 & 0.137 & $-0.017$ ($0.195$) & 0.874 ($0.197$) \\
\addlinespace[2pt]
12 & Weibull          & 0.112 & 0.148 & $-0.037$ ($0.001$) & 0.753 ($0.002$) \\
   & generalized gamma & 0.109 & 0.145 & $-0.037$ ($0.002$) & 0.746 ($0.002$) \\
\addlinespace[2pt]
24 & Weibull          & 0.107 & 0.148 & $-0.041$ ($<0.001$) & 0.723 ($<0.001$) \\
   & generalized gamma & 0.101 & 0.145 & $-0.044$ ($<0.001$) & 0.696 ($<0.001$) \\
\addlinespace[2pt]
36 & Weibull          & 0.105 & 0.144 & $-0.039$ ($<0.001$) & 0.728 ($<0.001$) \\
   & generalized gamma & 0.094 & 0.140 & $-0.046$ ($<0.001$) & 0.671 ($<0.001$) \\
\bottomrule
\end{tabular}
\end{table}

\section{Numerical studies under the second allocation pattern\label{app:sim}}
\setcounter{table}{0}\setcounter{figure}{0}

Tables~\ref{tab:sim-p2-cens1} and~\ref{tab:sim-p2-cens0} report the second allocation pattern: 70\% and 30\% of the sample in Strata~A and~B, 1:1 allocation within each stratum, and weights fixed at 0.7 and 0.3. All other settings are as in Section~\ref{ssec:sim-config-supp}. The findings were the same as in Section~\ref{sec:simulation}. The nonparametric procedure was never flagged at $\tau=6$ and was flagged in almost every replication at $\tau=24$ and $36$ months; at $\tau=12$ months, it was flagged in 302 and 61 of the 3,000 replications under random censoring ($n=600$ and $1{,}000$) and in 36 and 2 without censoring. Where it could be summarized with a sparse risk set (no censoring, $\tau=12$ months), its conditional coverage ranged from 0.991 to 1.000 for the control-group AH, the DAH, and $\log(\mathrm{RAH})$. Coverage ranged from 0.943 to 0.961 for the Weibull and from 0.942 to 0.961 for the generalized gamma estimator. The generalized gamma fit failed in more than 2\% of the replications in two configurations (random censoring, $n=600$, $\tau=24$ and $36$ months; 61 and 73 of 3,000), which are therefore not reported.

\begin{table}[!tbp]
\centering
\caption{Bias and coverage of the nonparametric and parametric procedures for the standardized AH, the DAH, and $\log(\mathrm{RAH})$, based on 3,000 replications under the second allocation pattern (70:30 stratum split, 1:1 allocation) and censoring pattern~II (no censoring). $\overline{m}_j(\tau)$ and the flag are defined in Section~\ref{ssec:sim-config-supp}. Nonparametric summaries are conditional on the absence of the flag; generalized gamma summaries are conditional on convergence.}
\label{tab:sim-p2-cens0}
\footnotesize
\setlength{\tabcolsep}{3pt}
\renewcommand{\arraystretch}{1.16}
\resizebox{\textwidth}{!}{%
\begin{tabular}{p{1.8cm}p{1.8cm}cccccccc}
\toprule
\multicolumn{2}{c}{sample size} & \multicolumn{4}{c}{N = 600} & \multicolumn{4}{c}{N = 1000} \\
\cmidrule(lr){3-6}\cmidrule(lr){7-10}
\multicolumn{2}{c}{tau} & 6 & 12 & 24 & 36 & 6 & 12 & 24 & 36 \\
\midrule
\multicolumn{10}{l}{\textbf{Average minimum positive risk-set size up to $\tau$, $\overline{m}_j(\tau)$}} \\
AH treatment & & 46.55 & 23.46 & 6.27 & 2.16 & 76.71 & 38.29 & 9.78 & 3.02 \\
AH control & & 50.27 & 5.21 & 1 & 1 & 83.4 & 8.12 & 1 & 1 \\
\addlinespace[2pt]
\multicolumn{10}{l}{\textbf{True value}} \\
AH treatment & & 0.09 & 0.095 & 0.098 & 0.099 & 0.09 & 0.095 & 0.098 & 0.099 \\
AH control & & 0.15 & 0.17 & 0.172 & 0.172 & 0.15 & 0.17 & 0.172 & 0.172 \\
DAH & & -0.059 & -0.076 & -0.074 & -0.073 & -0.059 & -0.076 & -0.074 & -0.073 \\
logRAH & & -0.506 & -0.589 & -0.564 & -0.555 & -0.506 & -0.589 & -0.564 & -0.555 \\
\addlinespace[2pt]
\rowcolor[gray]{0.9}
\multicolumn{2}{>{\cellcolor[gray]{0.9}}p{3.8cm}}{No. of replications with an empty risk set\newline at $\tau$ in some cell} & 0 & 36 & 2995 & 3000 & 0 & 2 & 2994 & 3000 \\
\multicolumn{2}{p{3.6cm}}{\# replications where the Weibull fit did not converge} & 0 & 0 & 0 & 0 & 0 & 0 & 0 & 0 \\
\multicolumn{2}{p{3.6cm}}{\# replications where the generalized gamma fit did not converge} & 1 & 0 & 2 & 2 & 0 & 0 & 0 & 0 \\
\midrule
\multicolumn{10}{l}{\textbf{Bias (MCSE) [$\times 10^{-3}$]}} \\
\multirow[t]{3}{*}{logRAH} & nonparametric & -0.32 (1.97) & 0.07 (1.49) & - & - & 1.34 (1.54) & 1.66 (1.16) & - & - \\
& Weibull & -0.38 (1.97) & -0.13 (1.48) & 0.30 (1.30) & 0.14 (1.24) & 1.46 (1.53) & 1.75 (1.16) & 1.95 (1.01) & 1.75 (0.98) \\
& GG & -0.02 (1.97) & -0.07 (1.48) & 0.41 (1.30) & 0.30 (1.24) & 1.62 (1.53) & 2.21 (1.16) & 1.98 (1.01) & 1.85 (0.98) \\
\multirow[t]{3}{*}{DAH} & nonparametric & 0.06 (0.22) & 0.01 (0.18) & - & - & 0.18 (0.18) & 0.19 (0.14) & - & - \\
& Weibull & 0.06 (0.22) & -0.01 (0.18) & -0.06 (0.17) & -0.10 (0.16) & 0.21 (0.18) & 0.21 (0.14) & 0.19 (0.13) & 0.15 (0.13) \\
& GG & 0.05 (0.22) & -0.01 (0.18) & -0.03 (0.17) & -0.05 (0.16) & 0.20 (0.18) & 0.31 (0.14) & 0.20 (0.13) & 0.18 (0.13) \\
\multirow[t]{3}{*}{AH treatment} & nonparametric & 0.26 (0.14) & 0.27 (0.12) & - & - & 0.25 (0.11) & 0.27 (0.09) & - & - \\
& Weibull & 0.25 (0.14) & 0.26 (0.12) & 0.30 (0.10) & 0.30 (0.10) & 0.25 (0.11) & 0.27 (0.09) & 0.27 (0.08) & 0.27 (0.07) \\
& GG & 0.33 (0.14) & 0.27 (0.12) & 0.29 (0.10) & 0.27 (0.10) & 0.29 (0.11) & 0.25 (0.09) & 0.27 (0.08) & 0.26 (0.07) \\
\multirow[t]{3}{*}{AH control} & nonparametric & 0.20 (0.17) & 0.26 (0.14) & - & - & 0.07 (0.14) & 0.08 (0.11) & - & - \\
& Weibull & 0.19 (0.17) & 0.27 (0.14) & 0.36 (0.13) & 0.40 (0.13) & 0.04 (0.14) & 0.06 (0.11) & 0.08 (0.11) & 0.12 (0.11) \\
& GG & 0.28 (0.17) & 0.27 (0.14) & 0.31 (0.13) & 0.32 (0.13) & 0.10 (0.14) & -0.06 (0.11) & 0.06 (0.11) & 0.08 (0.11) \\
\midrule
\multicolumn{10}{l}{\textbf{Coverage}} \\
\multirow[t]{3}{*}{logRAH} & nonparametric & 0.960 & 0.991 & - & - & 0.957 & 0.994 & - & - \\
& Weibull & 0.948 & 0.949 & 0.951 & 0.952 & 0.947 & 0.946 & 0.950 & 0.951 \\
& GG & 0.950 & 0.948 & 0.953 & 0.955 & 0.947 & 0.946 & 0.949 & 0.952 \\
\multirow[t]{3}{*}{DAH} & nonparametric & 0.965 & 0.998 & - & - & 0.962 & 0.998 & - & - \\
& Weibull & 0.953 & 0.953 & 0.961 & 0.961 & 0.943 & 0.944 & 0.949 & 0.950 \\
& GG & 0.954 & 0.953 & 0.961 & 0.961 & 0.943 & 0.943 & 0.948 & 0.948 \\
\multirow[t]{3}{*}{AH treatment} & nonparametric & 0.955 & 0.954 & - & - & 0.949 & 0.951 & - & - \\
& Weibull & 0.949 & 0.944 & 0.943 & 0.948 & 0.945 & 0.945 & 0.950 & 0.949 \\
& GG & 0.946 & 0.945 & 0.945 & 0.948 & 0.945 & 0.945 & 0.950 & 0.950 \\
\multirow[t]{3}{*}{AH control} & nonparametric & 0.969 & 1.000 & - & - & 0.960 & 1.000 & - & - \\
& Weibull & 0.948 & 0.952 & 0.955 & 0.955 & 0.943 & 0.945 & 0.947 & 0.949 \\
& GG & 0.951 & 0.952 & 0.955 & 0.953 & 0.942 & 0.946 & 0.947 & 0.947 \\
\bottomrule
\end{tabular}}
\end{table}

\begin{table}[!tbp]
\centering
\caption{Bias and coverage of the nonparametric and parametric procedures for the standardized AH, the DAH, and $\log(\mathrm{RAH})$, based on 3,000 replications under the second allocation pattern (70:30 stratum split, 1:1 allocation) and censoring pattern~I (random censoring). $\overline{m}_j(\tau)$ and the flag are defined in Section~\ref{ssec:sim-config-supp}. Nonparametric summaries are conditional on the absence of the flag; generalized gamma summaries are conditional on convergence.}
\label{tab:sim-p2-cens1}
\footnotesize
\setlength{\tabcolsep}{3pt}
\renewcommand{\arraystretch}{1.16}
\resizebox{\textwidth}{!}{%
\begin{tabular}{p{1.8cm}p{1.8cm}cccccccc}
\toprule
\multicolumn{2}{c}{sample size} & \multicolumn{4}{c}{N = 600} & \multicolumn{4}{c}{N = 1000} \\
\cmidrule(lr){3-6}\cmidrule(lr){7-10}
\multicolumn{2}{c}{tau} & 6 & 12 & 24 & 36 & 6 & 12 & 24 & 36 \\
\midrule
\multicolumn{10}{l}{\textbf{Average minimum positive risk-set size up to $\tau$, $\overline{m}_j(\tau)$}} \\
AH treatment & & 34.82 & 12.89 & 2.27 & 1.06 & 57.37 & 20.54 & 3.12 & 1.14 \\
AH control & & 37.05 & 3.25 & 1 & 1 & 61.95 & 4.74 & 1 & 1 \\
\addlinespace[2pt]
\multicolumn{10}{l}{\textbf{True value}} \\
AH treatment & & 0.09 & 0.095 & 0.098 & 0.099 & 0.09 & 0.095 & 0.098 & 0.099 \\
AH control & & 0.15 & 0.17 & 0.172 & 0.172 & 0.15 & 0.17 & 0.172 & 0.172 \\
DAH & & -0.059 & -0.076 & -0.074 & -0.073 & -0.059 & -0.076 & -0.074 & -0.073 \\
logRAH & & -0.506 & -0.589 & -0.564 & -0.555 & -0.506 & -0.589 & -0.564 & -0.555 \\
\addlinespace[2pt]
\rowcolor[gray]{0.9}
\multicolumn{2}{>{\cellcolor[gray]{0.9}}p{3.8cm}}{No. of replications with an empty risk set\newline at $\tau$ in some cell} & 0 & 302 & 2999 & 3000 & 0 & 61 & 2998 & 3000 \\
\multicolumn{2}{p{3.6cm}}{\# replications where the Weibull fit did not converge} & 0 & 0 & 0 & 0 & 0 & 0 & 0 & 0 \\
\multicolumn{2}{p{3.6cm}}{\# replications where the generalized gamma fit did not converge} & 1 & 0 & 61 & 73 & 1 & 0 & 10 & 10 \\
\midrule
\multicolumn{10}{l}{\textbf{Bias (MCSE) [$\times 10^{-3}$]}} \\
\multirow[t]{3}{*}{logRAH} & nonparametric & -2.44 (2.12) & - & - & - & -2.58 (1.66) & - & - & - \\
& Weibull & -2.19 (2.12) & -0.51 (1.67) & 0.46 (1.59) & 0.13 (1.58) & -2.49 (1.65) & -2.13 (1.31) & -1.47 (1.25) & -1.32 (1.23) \\
& GG & -2.31 (2.13) & -0.97 (1.68) & - & - & -2.69 (1.66) & -2.25 (1.32) & -1.34 (1.27) & -1.08 (1.27) \\
\multirow[t]{3}{*}{DAH} & nonparametric & -0.25 (0.24) & - & - & - & -0.18 (0.19) & - & - & - \\
& Weibull & -0.25 (0.24) & -0.05 (0.21) & -0.00 (0.20) & -0.03 (0.21) & -0.18 (0.19) & -0.20 (0.16) & -0.15 (0.16) & -0.13 (0.16) \\
& GG & -0.25 (0.24) & -0.06 (0.21) & - & - & -0.20 (0.19) & -0.20 (0.16) & -0.11 (0.16) & -0.04 (0.16) \\
\multirow[t]{3}{*}{AH treatment} & nonparametric & 0.30 (0.16) & - & - & - & -0.03 (0.12) & - & - & - \\
& Weibull & 0.36 (0.15) & 0.30 (0.13) & 0.38 (0.13) & 0.32 (0.13) & -0.01 (0.12) & -0.00 (0.10) & 0.02 (0.10) & 0.01 (0.10) \\
& GG & 0.33 (0.16) & 0.22 (0.13) & - & - & -0.01 (0.12) & -0.02 (0.10) & 0.01 (0.10) & -0.03 (0.10) \\
\multirow[t]{3}{*}{AH control} & nonparametric & 0.56 (0.19) & - & - & - & 0.15 (0.15) & - & - & - \\
& Weibull & 0.61 (0.19) & 0.35 (0.16) & 0.38 (0.16) & 0.34 (0.16) & 0.17 (0.14) & 0.20 (0.12) & 0.18 (0.13) & 0.14 (0.13) \\
& GG & 0.58 (0.19) & 0.29 (0.16) & - & - & 0.19 (0.15) & 0.17 (0.12) & 0.12 (0.13) & 0.02 (0.13) \\
\midrule
\multicolumn{10}{l}{\textbf{Coverage}} \\
\multirow[t]{3}{*}{logRAH} & nonparametric & 0.957 & - & - & - & 0.957 & - & - & - \\
& Weibull & 0.949 & 0.950 & 0.950 & 0.949 & 0.948 & 0.946 & 0.946 & 0.947 \\
& GG & 0.945 & 0.950 & - & - & 0.948 & 0.945 & 0.946 & 0.946 \\
\multirow[t]{3}{*}{DAH} & nonparametric & 0.961 & - & - & - & 0.959 & - & - & - \\
& Weibull & 0.945 & 0.947 & 0.947 & 0.947 & 0.949 & 0.950 & 0.944 & 0.944 \\
& GG & 0.947 & 0.951 & - & - & 0.949 & 0.948 & 0.946 & 0.942 \\
\multirow[t]{3}{*}{AH treatment} & nonparametric & 0.949 & - & - & - & 0.947 & - & - & - \\
& Weibull & 0.949 & 0.951 & 0.953 & 0.953 & 0.947 & 0.952 & 0.952 & 0.954 \\
& GG & 0.949 & 0.950 & - & - & 0.948 & 0.952 & 0.947 & 0.949 \\
\multirow[t]{3}{*}{AH control} & nonparametric & 0.972 & - & - & - & 0.966 & - & - & - \\
& Weibull & 0.953 & 0.947 & 0.945 & 0.948 & 0.946 & 0.950 & 0.945 & 0.944 \\
& GG & 0.949 & 0.946 & - & - & 0.948 & 0.953 & 0.947 & 0.950 \\
\bottomrule
\end{tabular}}
\end{table}


\bibliographystyle{vancouver}
\bibliography{refs}

\end{document}